\documentclass[lineno]{jfm}
\usepackage[applemac]{inputenc}
\usepackage[english]{babel}
\usepackage{listings} 
\usepackage{newtxtext}
\usepackage{newtxmath}
\usepackage{curve2e}
\usepackage{fancyhdr}
\usepackage{lscape}
\usepackage{graphicx}
\usepackage[caption=false]{subfig}
\usepackage{mathtools}
\usepackage{amsfonts}
\usepackage{acronym}
\usepackage{psfrag}
\usepackage{amssymb}
\usepackage{ifsym}
\usepackage{nomencl}
\usepackage{multirow}
\usepackage{multicol}
\usepackage{placeins}
\usepackage{algorithm}
\usepackage{algpseudocode}
\usepackage{color}
\usepackage{cases}
\usepackage{float}
\usepackage{subfloat}
\usepackage[breaklinks=true]{hyperref}
\hypersetup{
    colorlinks = true,
    urlcolor   = blue,
    citecolor  = black,
}

\newcommand{\RomanNumeralCaps}[1]
\linenumbers
\usepackage{enumerate}
\usepackage{rotating}
\usepackage{rotfloat}
\usepackage{tabulary}
\usepackage{natbib}
\usepackage{stmaryrd}
\usepackage{setspace}
\usepackage{verbatim}
\usepackage{breakcites}
\usepackage{ar}
\usepackage{tikz,tikz-3dplot}
\usepackage{url}

\usetikzlibrary{arrows,shapes,calc}
\usetikzlibrary{decorations.pathreplacing}

\newcommand\St{\mbox{\textit{St}}}  
\newcommand\Nu{\mbox{\textit{Nu}}} 
\newcommand\Ri{\mbox{\textit{Ri}}} 

\definecolor{darkviolet}{rgb}{0.58, 0.0, 0.83}
\definecolor{shamrockgreen}{rgb}{0.0, 0.62, 0.38}
\definecolor{deepskyblue}{rgb}{0.0, 0.75, 1.0}
\definecolor{amber}{rgb}{1.0, 0.49, 0.0}
\definecolor{red}{rgb}{1.0, 0.0, 0.0}

\newcommand\diff{\mbox{\textit{d}}}  

\title{Friction and heat transfer in forced air convection with variable physical properties}
\author{Davide Modesti$^{1}$\corresp{\email{d.modesti@tudelft.nl}} \and Sergio Pirozzoli$^2$}

\affiliation{\aff{1} Gran Sasso Science Institute, Viale Francesco Crispi 7, 67100, L'Aquila, Italy 
\aff{2} Dipartimento di Ingegneria Meccanica e Aerospaziale, Sapienza Universit\`a di Roma, via Eudossiana 18, 00184 Roma, Italia}

\begin{document}
\maketitle


%


\begin{abstract}
We establish a theoretical framework for predicting friction and heat transfer coefficients 
in variable-property forced air convection. Drawing from concepts 
in high-speed wall turbulence, which also involves significant temperature, viscosity, and density variations, 
we utilize the mean momentum balance and mean thermal balance equations to develop integral transformations 
that account for the impact of variable fluid properties. These transformations are then applied inversely to predict the friction and heat transfer coefficients, leveraging the universality of passive scalars transport theory.
Our proposed approach is validated using a comprehensive dataset from direct numerical simulations, 
covering both heating and cooling conditions up to a friction Reynolds number 
of approximately $\Rey_\tau\approx 3200$. The predicted friction and heat transfer coefficients 
closely match DNS data with an accuracy margin of 1-2\%, 
representing a significant improvement over the current state of the art.
\end{abstract}


\section{Introduction}

Heat transfer by turbulent forced convection occurs when a cold fluid flows over a hot wall, or vice versa. 
Forced thermal convection has countless applications in engineering, and it is the fundamental principle upon which heat exchangers are designed and built~\citep{incropera_96,kakac_02}. Heat exchangers are a key component in any energy conversion system,
for instance, the radiators in our homes, heat pumps, fuel cells, nuclear plants, and solar receivers. 
In aerospace engineering, two notable applications are aircraft and rocket engines, where components are subjected to extreme heat loads and internal cooling is necessary to guarantee the material integrity.

Most studies on forced thermal convection regard the temperature field as a passive scalar, neglecting its feedback effect on the velocity field through the variation of the transport properties of the fluid. 
Notable experimental and numerical studies relying on the constant-property assumption are the ones by~\citet{sparrow_66,xia_22,alcantara_21,abe_17}, 
together with the more recent direct numerical simulation (DNS) studies performed by our group~\citep{pirozzoli_16,pirozzoli_23}.

The constant-property assumption is valid if temperature variations are on the order of a few percent because the thermodynamic variables and fluid properties
can be assumed to be constant~\citep{cebeci_84}. However, most engineering applications feature large temperature differences. For instance, in aircraft engines, 
the air passing through the cooling channels of turbine blades has a mean temperature $T_m \approx 400$K, whereas the wall temperature reaches $T_w\approx800$K, 
hence their ratio is well beyond the range of validity of the constant-property assumption. 
This range of temperature variations is common in engineering applications, 
however the constant-property assumption is used in most academic research and exploited in the engineering practice. 

Preliminary design of cooling/heating ducts is based mainly on predictive formulas for the Nusselt number and the pressure drop, which are used for sizing the cooling passages. 
Among the most classical engineering formulas, we recall those by Dittus--Boelter~\citep{dittus_33} and Gnielinski~\citep{gnielinski_75} for the Nusselt number, 
and Prandtl's friction formula for the pressure drop~\citep{nikuradse_33}. Although their use is widespread in engineering design, 
these formulas are based on the constant-property assumption and do not directly account for the effect of variable fluid properties, which is usually included using empirical corrections~\citep{yeh_84,sleicher_75}.

The most popular empirical corrections for heat transfer prediction in water are those by~\citet{dittus_85} and~\citet{sieder_36}, 
which account for the fluid viscosity variations through an empirical corrective factor $\left(\mu_b/\mu_w\right)^n$ applied to the Nusselt number resulting from formulas obtained for the constant-property case, 
where $\mu_m$ and $\mu_w$ are the viscosities of the fluid evaluated at the mean and wall temperatures, respectively.
Also for gases, many empirical predictive formulas for the Nusselt number are available, and they have been reviewed extensively by~\citet{petukhov_70,yeh_84}. 
However, they all have a similar structure to those used for water, relying on a correction factor based on the mean-to-wall temperature ratio $\left(T_m/T_w\right)^n$, 
with exponent $n$ depending on the type of gas and on the cooling/heating ratio. 
Similar corrections are also used to estimate the friction factor and they suggest drag reduction in the case of wall heating, both for liquids~\citep{sieder_36} and gases~\citep{yeh_84}, compared to the adiabatic case.
However, these corrections are fluid-dependent, are available only for a limited number of fluids, and their accuracy is often questionable.

More recently, some authors have studied forced thermal convection in fluids with variable properties using DNS. 
\citet{zonta_12} performed DNS of water flow in a plane channel with a heated and a cold wall and found a reduction of the Reynolds shear stress and 
of the friction coefficient at the heated wall, 
despite the lower viscosity which increases the local Reynolds number.
\citet{lee_13} performed DNS of turbulent boundary layers with temperature-dependent viscosity representative of water and investigated the effect of wall heating 
on the friction and heat transfer coefficients. They reported a 26\% drag reduction for water with a freestream-to-wall temperature ratio of 0.77, at a freestream temperature of about $300$K, 
and the drag reduction mechanism was attributed to a reduction of the wall-shear stress, in agreement with the findings of~\citet{zonta_12}. 
\citet{lee_14} used the same DNS dataset to assess heat transfer modifications due to variable viscosity effects and proposed a 
correction to the classical Kader's fitting for the mean temperature profile~\citep{kader_81}.
\citet{patel_16,patel_17} studied the effects of variable density and viscosity in liquid-like and gas-like fluids using DNS. They utilized compressibility transformations, 
originally developed for high-speed boundary layers~\citep{modesti_16,trettel_16}, to map velocity and temperature profiles to the constant-property case, reporting
good agreement with the constant-property profiles.
\citet{kaller_19} conducted a wall-resolved large-eddy simulation of flow in a duct with one heated side, filled with water, 
and observed reduced friction near the heated wall, which was also accompanied by weakened secondary flows. 
The effect of density variations is also important in the context of mixed convection,
although the common practice is to rely on the Boussinesq approximation~\citep{pinelli_10,yerragolam_24},
whereas studies that account for non-Oberbeck--Boussinesq effect are more limited~\citep{zonta_13}.
The effect of variable physical properties is also important in pure natural convection~\citep{gray_76}, particularly in experimental studies, where it is challenging to achieve high Rayleigh numbers while satisfying the Oberbeck--Boussinesq approximation.
To the best of our knowledge, there is no counterpart of the Grossman--Lohse theory~\citep{grossmann_00} for Rayleigh--B\'enard convection with variable properties.

Although studies focusing on the effect of density and viscosity variations in forced thermal convection are available, predictive formulas for the heat transfer and 
friction coefficients are invariably based on empirical fitting of experimental data, and the few numerical studies available did not discuss in detail the prediction of these coefficients.
In this study, we aim to develop a more solid theoretical framework to estimate the mean friction drag and heat transfer in the presence of variation of the transport properties, focusing on the case of air as working fluid. 
For that purpose, we use DNS data of plane turbulent channel flow at a moderate Reynolds number to develop improved formulas for friction
and heat transfer prediction.

\section{Methodology}\label{sec:method}
We solve the compressible Navier--Stokes equations using our flow solver STREAmS~\citep{bernardini_21,bernardini_23},
additional details on the numerical method are reported in the Appendix~\ref{appA}.
The streamwise momentum equation is forced in such a way as to maintain a constant mass flow rate. 
Periodicity is exploited in the streamwise and spanwise directions, and isothermal no-slip boundary conditions are used at the channel walls. Let $h$ be the half-width of the channel, 
the DNS have been carried out in a computational domain $L_x\times L_y\times L_z=6\pi h \times 2 h\times 2\pi h$. 
A uniform bulk cooling or heating term is added to the entropy equation to guarantee that the mixed mean temperature, defined as,
\begin{equation}
	T_m = \frac{1}{2h \rho_b u_b}\int_{0}^{2h}\overline{\rho}\, \overline{u} \overline{T} \mathrm{d}y,\quad \rho_b=\frac{1}{2h}\int_{0}^{2h}\overline{\rho}\mathrm{d}y,\quad u_b = \frac{1}{2h \rho_b}\int_{0}^{2h}\overline{\rho}\, \overline{u} \mathrm{d}y,
	\label{eq:bulk}
\end{equation}
remains exactly constant in time. Here, $\rho_b$ and $u_b$ are the bulk density and velocity, respectively.
In the following, the overline symbol is used to indicate Reynolds averaging in time and in the homogeneous spatial directions, and the prime superscript is used to denote fluctuations thereof.
As common in variable-density flows, we also use Favre averages, denoted with the tilde superscript $\widetilde{f}=\overline{\rho f}/\overline{\rho}$, 
and the double prime superscript will indicate fluctuations thereof.

The $+$ superscript is used to denote normalization by wall units, namely by friction velocity, 
$u_{\tau} = (\tau_w/\overline{\rho}_w)^{1/2}$ (where $\tau_w= \mu_w {\diff \widetilde{u}}/{\diff y}\vert_w$ is the mean wall shear stress), and the associated viscous length scale, $\delta_v=\nu_w/u_\tau$,
where the subscript 'w' denotes quantities evaluated at the wall.
For inner normalization of the mean temperature, we use the friction temperature, $\theta_{\tau} = q_w/(\rho_w c_p u_{\tau})$, where $q_w = \lambda_w {\diff \widetilde{T}}/{\diff y}\vert_w$ is the mean wall heat flux, 
where $c_p=\gamma/(\gamma-1) R$ is the specific heat capacity at constant pressure, 
$R$ the air constant, and $\lambda_w$ is the thermal conductivity at the wall,
evaluated as $\lambda = \mu c_p/\Pran$, with Prandtl number set to $\Pran=0.72$.

\begin{landscape}
\begin{table}
	\centering
	\begin{tabular}{lrrrrrrrrrrrrlr}
\hline
                 &   $\Rey_b$ &   $\Rey_\tau$ &   $T_m/T_w$ &   $T_w(K)$ &   ${\Rey_{\tau}}_{cp}$ &   $C_f\times10^3$ &   $\St\times10^3$ &   $Nu$ &   $N_x$ &   $N_y$ &   $N_z$ &   ${\Delta x^*}$ & $\Delta y^*_w$--$\Delta y^*_{max}$   &   $\Delta z^*$ \\
\hline
 L04             &      17182 &           155 &         0.4 &     800    &             533 &              5.45 &              3.09 &   38.3 &    1024 &     280 &     512 &            9.8 & 0.69--5.1                            &            6.5 \\
 L05-A           &      20170 &           212 &         0.5 &     293.15 &             612 &              5.57 &              3.18 &   46.2 &    1024 &     280 &     512 &           11.3 & 0.67--5.9                            &            7.5 \\
 L05-B      &      17115 &           205 &         0.5 &     800    &             511 &              5.62 &              3.21 &   39.5 &    1024 &     280 &     512 &            9.4 & 0.58--4.9                            &            6.3 \\
 L07             &      13565 &           255 &         0.7 &     800    &             400 &              6.28 &              3.56 &   34.8 &    1024 &     280 &     512 &            7.4 & 0.40--3.8                            &            4.9 \\
 L08             &      16679 &           356 &         0.8 &     800    &             470 &              6.01 &              3.4  &   40.9 &    1024 &     280 &     512 &            8.7 & 0.47--4.5                            &            5.8 \\
 L15             &      14632 &           663 &         1.5 &     800    &             406 &              6.82 &              4.1  &   43.2 &    1024 &     280 &     512 &            7.5 & 0.41--3.9                            &            5   \\
 L2              &      11389 &           789 &         2   &     293.15 &             317 &              7.31 &              4.37 &   35.9 &    1024 &     280 &     512 &            5.8 & 0.32--3.0                            &            3.9 \\
 L25             &       9853 &           902 &         2.5 &     293.15 &             278 &              7.75 &              4.62 &   32.8 &    1024 &     280 &     512 &            5.1 & 0.28--2.7                            &            3.4 \\
 L3              &       9212 &          1051 &         3   &     293.15 &             261 &              8.04 &              4.78 &   31.7 &    1024 &     280 &     512 &            4.8 & 0.18--2.5                            &            3.2 \\
 H04             &      31797 &           260 &         0.4 &     800    &             893 &              4.52 &              2.59 &   59.3 &    2048 &     480 &    1024 &            8.2 & 0.66--5.0                            &            5.5 \\
 H07             &      37887 &           617 &         0.7 &     800    &             971 &              4.75 &              2.73 &   74.5 &    2048 &     480 &    1024 &            8.9 & 0.45--5.4                            &            6   \\
 H05-A           &      37589 &           360 &         0.5 &     293.15 &            1040 &              4.66 &              2.68 &   72.5 &    2048 &     480 &    1024 &            9.6 & 0.37--5.8                            &            6.4 \\
 H05-B           &      37933 &           404 &         0.5 &     800    &            1007 &              4.47 &              2.57 &   70.3 &    2048 &     480 &    1024 &            9.3 & 0.57--5.6                            &            6.2 \\
 H08             &      34703 &           674 &         0.8 &     800    &             890 &              4.97 &              2.83 &   70.7 &    2048 &     480 &    1024 &            8.2 & 0.41--5.0                            &            5.5 \\
 H15             &      18694 &           861 &         1.5 &     293.15 &             498 &              6.3  &              3.78 &   50.8 &    2048 &     480 &    1024 &            4.6 & 0.27--2.8                            &            3.1 \\
 H2              &      15362 &          1028 &         2   &     293.15 &             415 &              6.81 &              4.06 &   44.9 &    2048 &     480 &    1024 &            3.8 & 0.15--2.3                            &            2.5 \\
 H25             &      13873 &          1224 &         2.5 &     293.15 &             378 &              7.15 &              4.26 &   42.5 &    2048 &     480 &    1024 &            3.5 & 0.13--2.1                            &            2.3 \\
 H3              &      12898 &          1420 &         3   &     293.15 &             354 &              7.44 &              4.42 &   41.1 &    2048 &     480 &    1024 &            3.3 & 0.12--2.0                            &            2.2 \\
 VH05          &      68874 &           680 &         0.5 &     800    &            1687 &              3.85 &              2.22 &  110.3 &    4096 &     800 &    2048 &            7.8 & 0.30--5.6                            &            5.2 \\
 VH2           &      54439 &          3201 &         2   &     293.15 &            1298 &              5.22 &              3.13 &  122.9 &    4096 &     800 &    2048 &            6   & 0.23--4.3                            &            4   \\
\hline
\end{tabular}
\caption{Flow parameters for plane channel flow DNS.
Box dimensions are $6\pi h \times 2h \times 2\pi h$ for all flow cases.
$\Rey_b = 2 \rho_b h u_b / \nu_m$ is the bulk Reynolds number and
$\Rey_{\tau} = h u_{\tau} / \nu_w$ is the friction Reynolds number.
${\Rey_{\tau}}_{cp} = y_{cp}(h)/\delta_v$ is the equivalent friction Reynolds number, defined in equation~\label{eq:Retaucp}
$T_m$ and $T_w$ and the mixed mean temperature and the wall temperature, respectively.
$C_f = 2 \tau_w /(\rho_b u_b^2)$ is the friction coefficient,
$\St=q_w/\left[\rho_b C_p u_b (T_w-T_m)\right]$ is the Stanton number and $\Nu=\St\Rey_b\Pran$ is the Nusselt number.
$\Delta x$ and $\Delta z$ are the mesh spacing in the streamwise and spanwise directions, and
	$\Delta y_w$ is the mesh spacing at the wall, where the $^*$ superscript indicates normalization with equivalent constant-property viscous length scale $\delta_{v,cp}$, 
defined in equation~\eqref{eq:retau_inc}.}
\label{tab:dataset}
\end{table}
\end{landscape}

Twenty DNS have been carried out at bulk Mach number $M_b=u_b/c_m=0.2$ (where $c_m$ is the speed of sound at the mixed mean temperature), 
and bulk Reynolds number $\Rey_b=2\rho_b u_b h/\mu_m\approx9000$--$70000$ (see table~\ref{tab:dataset}), where $\mu_m = \mu (T_m)$ is the dynamic viscosity evaluated at the mixed mean temperature, as obtained from Sutherland's law~\citep{white_74}.
The Mach number is low enough that compressibility effects are negligible, as it turns out, in order to isolate variable-property effects.
We consider various mean-to-wall 
temperature ratios, namely $T_m/T_w=\left[0.4,0.5,0.7,0.8,1.5,2,2.5,3\right]$, 
resulting in friction Reynolds numbers in the range $\Rey_\tau=u_\tau h/\nu_w\approx 150$--$3200$, where $\nu_w$ is the kinematic viscosity at the wall. 
For the case of mean-to-wall temperature ratio $T_m/T_w=0.5$, we also study the effect of 
varying the dimensional wall temperature, considering cases with $T_w=800K$
and $T_w=293.15K$.
This temperature range is rather wide and cover most applications of forced air convection we are aware of. 
Cases with wall heating approach the condensation temperature of air ($\approx 90$K) and cases with wall cooling
feature temperature variations in the range ($220$K $\lesssim T \lesssim 1000$ K), involving temperature values
beyond the ones normally found in heat exchangers.
For each value of the mean-to-wall temperature ratio, we have two flow cases denoted with
the letter L or H, depending on whether the Reynolds number is comparatively `low' or `high'.
For two flow cases with mean-to-wall temperature ratio $T_m/T_w=0.5$ and $T_m/T_w=2$ we also consider a `very high' Reynolds number case, denoted with VH.
In the present study, we consider pure forced convection neglecting the effect of buoyancy because in forced air convection
applications the Richardson number is typically very small. 
For instance, cooling channels of turbine blades have a bulk velocity $u_b\approx 30$--$60$m/s
and temperature variations of order $400$K, leading to Richardson numbers in the order $\Ri\approx 10^{-5}$.

\section{Instantaneous temperature field}

\begin{figure}
 \begin{center}
\includegraphics[width=12.cm,clip]{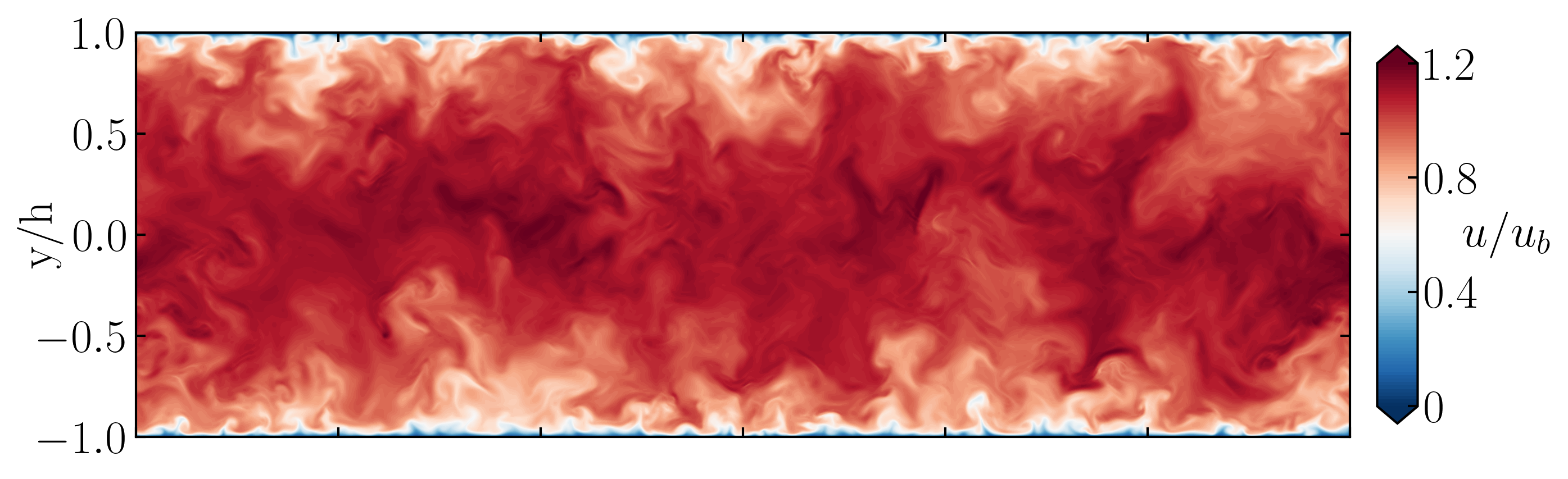}
\put(-340,100){(\textit{a})}\\
\includegraphics[width=12.cm,clip]{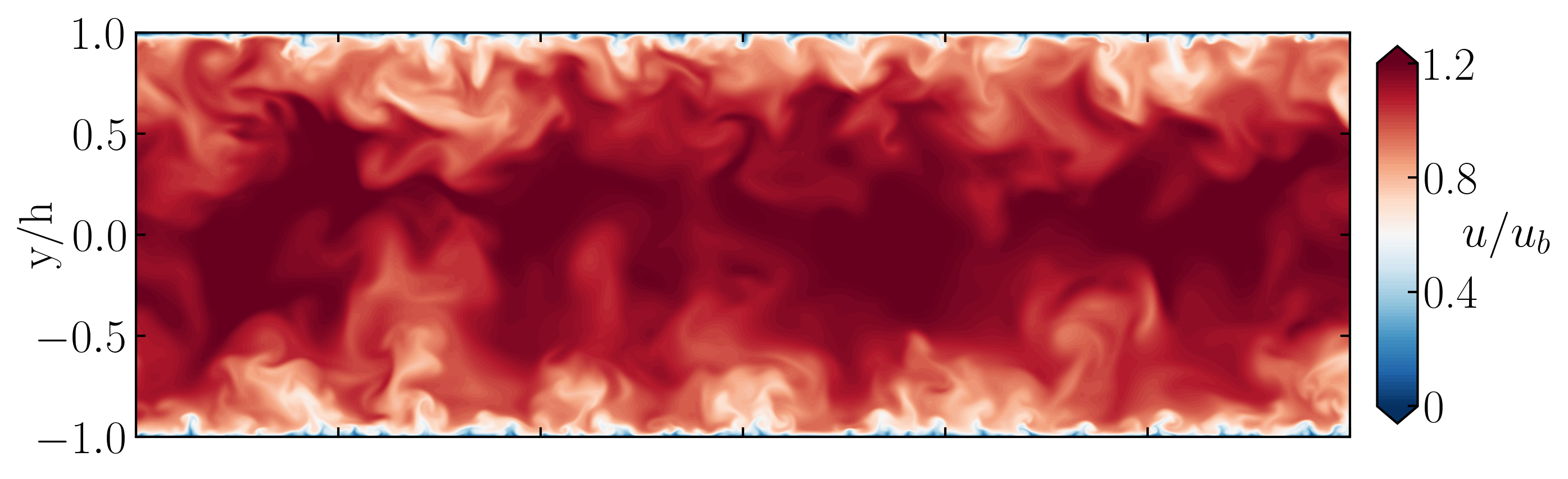}
\put(-340,100){(\textit{b})}\\
\includegraphics[width=12.cm,clip]{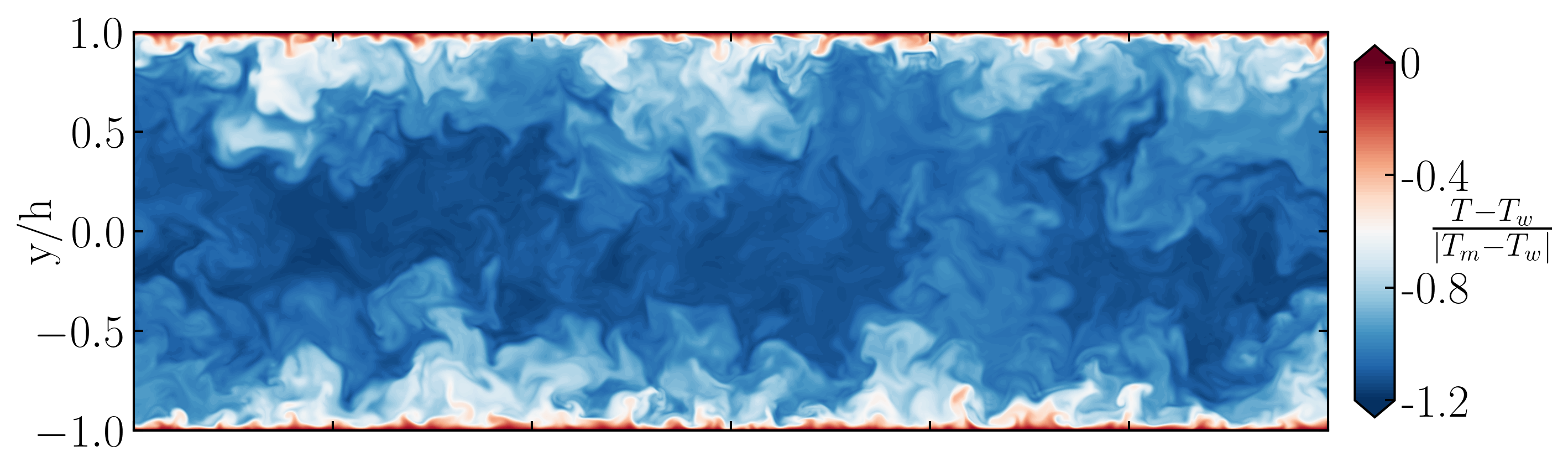}
\put(-340,100){(\textit{c})}\\
\includegraphics[width=12.cm,clip]{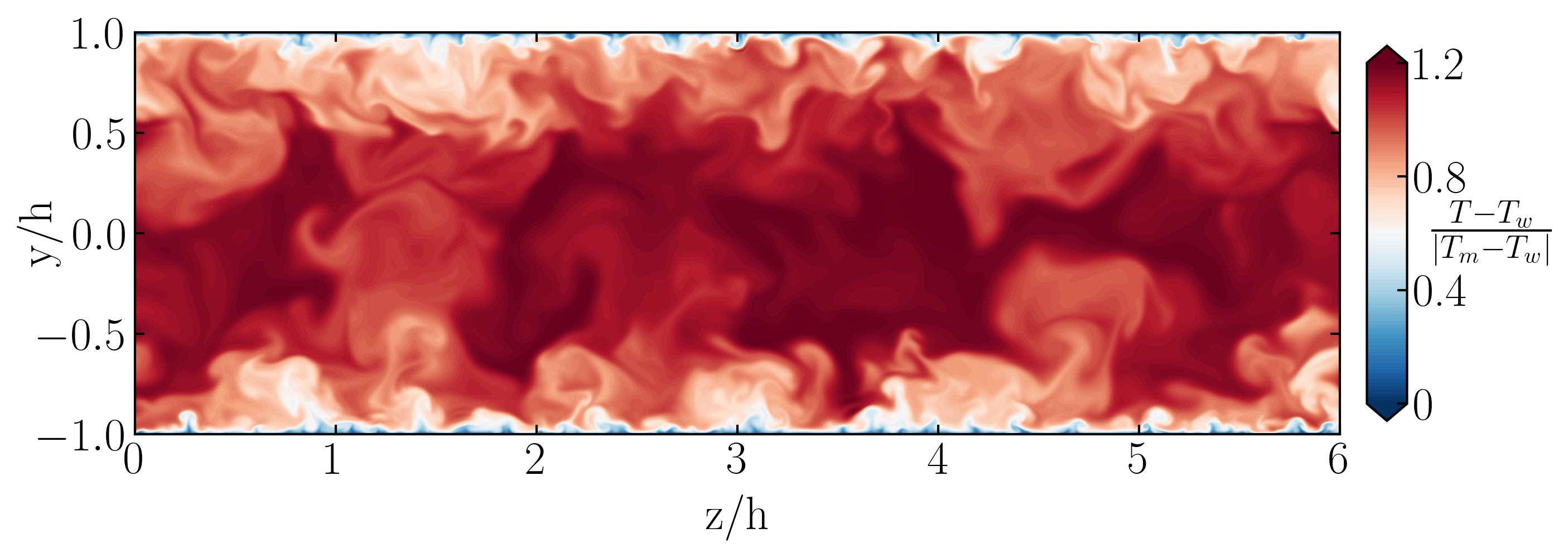}
\put(-340,115){(\textit{d})}\\
	 \caption{Instantaneous velocity (\textit{a},\textit{b}) and temperature (\textit{c},\textit{d}) fields in a cross-stream plane, 
	 for flow case H05-A (wall heating, $\Rey_\tau=360$, $T_m/T_w=0.5$) (\textit{a},\textit{c})
	 and H3 wall cooling ($\Rey_\tau=1420$, $T_m/T_w=3$) (\textit{b},\textit{d}).}
  \label{fig:inst}
 \end{center}
\end{figure}

We begin our analysis by inspecting the instantaneous velocity and temperature fields of flow cases H05-A (wall heating) and H3 (wall cooling) in figure~\ref{fig:inst}. 
Both cases exhibit the qualitative features that characterize wall turbulence, with high-speed cold (or hot) flow structures protruding towards the walls, 
and low-speed hot (or cold) fluid regions protruding towards the channel center. 
Despite sharing the general features of wall turbulence, we also note a significant effect of the thermodynamic and fluid property variations between cases with wall heating and wall cooling.
First, we observe that the friction Reynolds number values reported in table~\ref{tab:dataset} are not indicative of actual separation of 
scales in constant-property flows. For instance,
flow case with wall heating H05-A ($T_m/T_w=0.5$) has a relatively low friction Reynolds number ($\Rey_\tau=360$) but it exhibits finer eddies
than one would expect at this Reynolds number value.
This effect can be traced to strong viscosity variations within the near-wall and core flow regions. 
The opposite is true for wall cooling and, for instance, flow case H3 ($T_m/T_w=3$) has a higher friction Reynolds number ($\Rey_\tau=1420$) but small scales are absent and 
the flow appears to be a `low-pass filtered' version of the heated case, in which only large structures survive. 
In fact, in flow case H3 we observe large structures extending from one wall to beyond the channel centreline, 
whereas those are masked by smaller eddies in flow case H05-A.

We note that the instantaneous velocity and temperature fields are highly correlated, which is to be expected due to the similarity of
the underlying equations and the near-unity value of the Prandtl number, hence Reynolds analogy qualitatively holds also for the case
of variable fluid properties. Close scrutiny of the kinematic
and thermal fields reveals that temperature has finer structures as compared to velocity, which is partly
due to Prandtl number being lower than unity and partly to the effect of the pressure gradient term 
in the momentum equation~\ref{eq:momentum}, which is absent in the energy equation~\ref{eq:entropy} \citep{pirozzoli_16}.
The occurrence of sharper eddy boundaries in passive scalars compared to the velocity field, even at unit Prandtl number,
is a well known feature that has been reported by several authors~\citep{guezennec_90,abe_17}, and it has been associated to the
unmixedness of the scalar, as the absence of the pressure gradient results in reduced heat transport as compared to momentum transport.

\section{Mean flow field and variable-property transformations}

\begin{figure}
 \begin{center}
 \includegraphics[scale=1,clip]{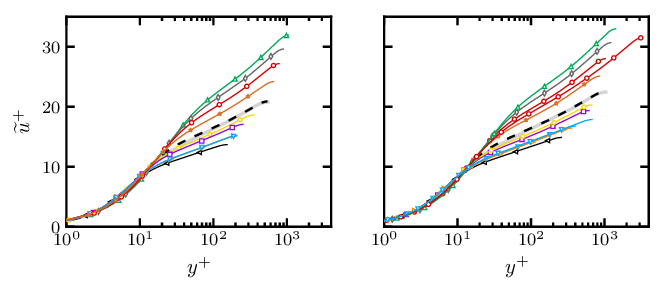}
  \put(-310,130){(\textit{a})}
  \put(-150,130){(\textit{b})}\\
 \includegraphics[scale=1,clip]{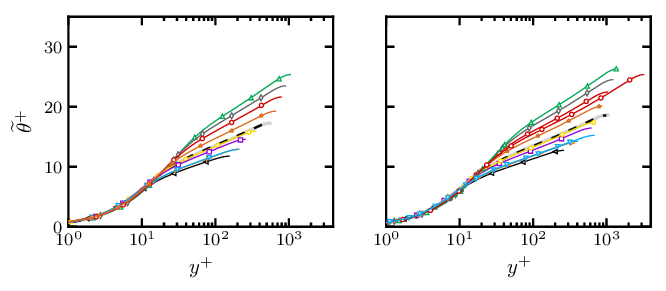}
  \put(-310,130){(\textit{c})}
  \put(-150,130){(\textit{d})}\\
	 \caption{Mean velocity (\textit{a},\textit{b}) and mean temperature (\textit{c},\textit{d}) profiles
	 for L flow cases (\textit{a},\textit{c}) and H flow cases (\textit{b},\textit{d}).
	 Symbols indicate DNS data for different mean-to-wall
	 temperature ratios: 
	 $T_m /T_w = 0.4$ (left triangles), 
	 $T_m /T_w = 0.5, T_w=800K$ (downward triangle),
	 $T_m /T_w = 0.5, T_w=273.25K$ (right triangles), 
	 $T_m /T_w = 0.7$ (squares), 
	 $T_m /T_w = 0.8$ (hexagons), 
	 $T_m /T_w = 1.5$ (stars), 
	 $T_m /T_w = 2$   (circles), 
	 $T_m /T_w = 2.5$ (diamonds), 
	 $T_m /T_w = 3$ (upward triangles).
	 Gray solid lines indicate the mean velocity and temperature profiles of the constant-property case at $\Pran=0.72$,
	 obtained using the composite profiles of~\citet{pirozzoli_24}. The dashed black lines indicate DNS of constant-property channel flow from~\citet{pirozzoli_16} at $\Pran=0.71$. 
	 }
  \label{fig:mean}
 \end{center}
\end{figure}

We begin the mean flow analysis by comparing the mean velocity and temperature profiles to the equivalent statistics for the constant-property case. 
For that purpose we rely both on DNS data of constant-property plane channel flow with passive scalar transport at $\Pran=0.71$ from~\citet{pirozzoli_16},
and on synthetic composite profiles, which are obtained by matching inner-layer velocity and temperature profiles with the corresponding outer-layer distributions.
The inner-layer profiles are obtained by integrating the eddy viscosity of~\citet{musker_79} for the velocity and the eddy diffusivity proposed by~\citet{pirozzoli_23a} for the temperature profile. In the outer layer we use Clauser's hypothesis of uniform eddy viscosity~\citep{clauser_56} and uniform eddy diffusivity~\citep{pirozzoli_16}.
A complete derivation of the composite profiles is available in~\citet{pirozzoli_24}.
Figure~\ref{fig:mean} shows that the composite profiles of mean temperature and velocity for the constant-property case are essentially indistinguishable from the DNS data, with the advantage that
the synthetic profiles are available at any Reynolds and Prandtl number.

We note that the statistics of the variable-property DNS are substantially different from the constant-property case when scaled in classical wall units. 
All the flow cases exhibit deviations from the reference, starting from the buffer region, and becoming  
more evident in the logarithmic region, where both the logarithmic slope and the additive constant deviate from the constant-property references.
Hence, we conclude that in the variable-property case the law-of-the-wall for both the mean velocity and temperature is not universal, but rather depends on the specific mean-to-wall-temperature ratio.

In analogy with what done in compressible flows, we assume that the effects of density and viscosity variations can be accounted for using suitable convolution integrals~\citep{modesti_16},
\begin{equation}
	y_{cp}=\int_0^y f_{cp} \, {\mathrm d}y, \quad
	u_{cp}=\int_0^{\tilde{u}} g_{cp} \, {\mathrm d}\tilde{u}, \quad \theta_{cp}=\int_0^{\tilde{\theta}} h_{cp} \, {\mathrm d}\tilde{\theta},
	\label{eq:transf}
\end{equation}
with kernel functions $f_{cp}$, $g_{cp}$, $h_{cp}$ to be specified such that the flow properties are mapped 
to the universal, constant-property case, denoted with the `cp' subscript.
In order to account for variable-property effect, we consider the streamwise mean momentum balance equation,
\begin{equation}
	\overline{\mu'\frac{\mathrm{d} u'}{\mathrm{d} y}} +  \overline{\mu}\frac{\mathrm{d} \widetilde{u}}{\mathrm{d} y}
        - \bar{\rho}\widetilde{u''v''} =
        \overline{\rho}_w u_{\tau}^2 \left(1-\eta\right),
	\label{eq:mmb}
\end{equation}
where $\eta=y/h$. 
Following~\citet{hasan_23}, we then introduce 
an eddy viscosity for the turbulent shear stress, such that
$-\overline{\rho} \widetilde{u''v''} = \overline{\rho}\nu_T\mathrm{d}\tilde{u}/\mathrm{d}y$.
Substituting the transformations~\eqref{eq:transf} into the streamwise mean momentum balance equation, and assuming $\overline{\mu'\mathrm{d} u'/\mathrm{d} y}\approx0$, one obtains
\begin{equation}
	\frac{\overline{\mu}}{\mu_w}\frac{f_{cp}}{g_{cp}}\left(1 + \frac{\nu_T}{\nu}\right)\frac{\mathrm{d}u_{cp}^+}{\mathrm{d}y_{cp}^+} = 1-\eta.
	\label{eq:mmb_1}
\end{equation}
Comparing equation~\eqref{eq:mmb_1} with the constant-property counterpart, 
\begin{equation}
 \left(1 + \frac{\nu_{T,cp}}{\nu_{cp}}\right)\frac{\mathrm{d}u_{cp}^+}{\mathrm{d}y_{cp}^+} = 1-\eta,
\end{equation}
we find the following relation between the two kernel functions $f_{cp}$ and $g_{cp}$,
\begin{equation}
 \frac{\overline{\mu}}{\mu_w}\frac{f_{cp}}{g_{cp}}\left(1+\frac{\nu_T}{\nu}\right)=1+\frac{\nu_{T,cp}}{\nu_{cp}}.
\end{equation}
A second condition is needed, which we find by enforcing that van Driest scaling~\citep{vandriest_51} holds in the logarithmic region,
as done by~\citet{trettel_16}. With this condition we find,
\begin{equation}
	f_{cp}=\frac {\mathrm{d}}{\mathrm{d} y} \left( \frac y{R^{1/2} N} \right),\quad g_{cp}=\left(\frac{1 + \nu_T/\nu}{1+\nu_{T,cp}/\nu}\right)R N \frac {\mathrm{d}}{\mathrm{d} y} \left( \frac y{R^{1/2} N} \right),
\label{eq:TL}
\end{equation}
where $N=\overline{\nu}/\nu_w$, $ R=\overline{\rho}/\overline{\rho}_w$.
The kernel functions~\eqref{eq:TL} are formally equivalent to the velocity transformation derived by~\citet{hasan_23}
for high-speed turbulent boundary layer, with eddy viscosities to be specified.
Using the model eddy viscosity of \citet{musker_79} for the baseline case 
of constant-property flow, herein we extend the model to account for 
variable-property effects by including an ad-hoc correction
depending on the mean-to-wall temperature ratio
\begin{equation}
	\displaystyle\frac{\nu_{T,cp}}{\nu} = \frac{\left(\kappa y_{cp}^+\right)^3}{\left(\kappa y_{cp}^+\right)^2 + C_{v1}^2},\quad \displaystyle\frac{\nu_T}{\nu} = \frac{\left(\kappa y_{cp}^+\right)^3}{\left(\kappa y_{cp}^+\right)^2 + C_{v1}^2 + \varphi(T_m/T_w)},
\end{equation}
where $\kappa=0.387$ is the assumed K\'arm\'an constant, and $C_{v1}=7.3$. 
Fitting the present DNS data (only flow cases at 'high Reynolds numbers', denoted as H, have been taken into account)
we have determined empirically the following expressions for the additive function $\varphi$,
\begin{equation}
 \varphi(T_m/T_w) = 
\begin{cases}
	-32\log{(T_m/T_w)} - 59(1-T_m/T_w)^2, \quad  & T_m/T_w<1\\ 
	5.6(1-T_m/T_w), \quad & T_m/T_w>1.
\end{cases}
\label{eq:ffunc}
\end{equation}

\begin{figure}
 \begin{center}
 \includegraphics[scale=1.]{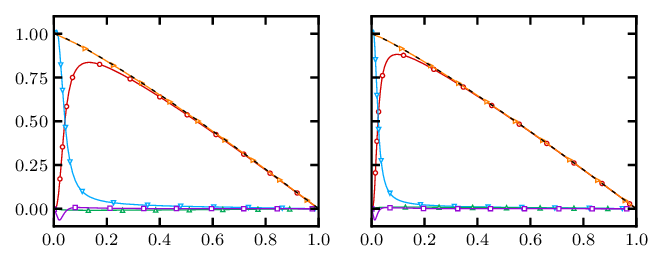}\\
 \includegraphics[scale=1.]{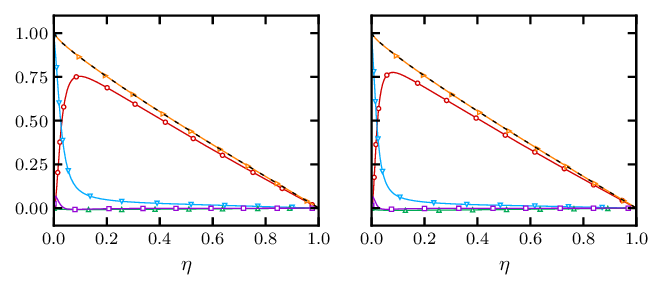}
 \caption{Mean energy balance as in equation~\eqref{eq:temp_balance}
	 for flow cases L05-A ($\Rey_\tau=212$, $T_m/T_w=0.5$) (\textit{a}), H05-A ($\Rey_\tau=360$, $T_m/T_w=0.5$) (\textit{b}), L3 ($\Rey_\tau=1051$, $T_m/T_w=3$) (\textit{c}), 
	 and H3 ($\Rey_\tau=1420$, $T_m/T_w=3$) (\textit{d}).
	 The symbols indicate mean conduction (downward triangles),
	 fluctuating conduction (squares), turbulent convection (circles),
	 dissipation (upward triangles), total heat flux $\mathcal{R\left({\eta}\right)}$ in equation
	 ~\eqref{eq:temp_balance}(dashed black line),
	 and sum of the different contributions (right triangles).
	 }
  \label{fig:temp_balance}
 \end{center}
\end{figure}

Similarly to what done for the mean velocity, a transformation for the mean temperature profile is 
obtained starting from the mean energy balance equation,
\begin{equation}
	\underbrace{\overline{\lambda'\frac{\mathrm{d} \theta'}{\mathrm{d} y}}}_{\text{Fluctuating Conduction}}
       +\underbrace{\overline{\lambda}\frac{\mathrm{d} \widetilde{\theta}}{\mathrm{d} y}}_{\text{Mean Conduction}}
	\underbrace{- C_p\bar{\rho}\,\widetilde{\theta''v''}}_{\text{Turbulent Convection}} + \underbrace{\Psi}_{\text{Dissipation}} =
	\underbrace{q_w \left(1-\mathcal{R}\right)}_{\text{Total heat flux}},
	\label{eq:temp_balance}
\end{equation}
where
\begin{equation}
\mathcal{R}(\eta) =\frac{1}{\rho_b} \int_0^\eta \overline{\rho}(\eta)\,\mathrm{d}\eta, \quad \Psi = \int_0^y \epsilon \, \mathrm{d}y -\int_0^y \frac{\rho}{\rho_b}\int_0^1\epsilon \, \mathrm{d}\eta\,\mathrm{d}y, 
\quad	\epsilon = \displaystyle\overline{\sigma_{ij}\frac{\partial u_i}{\partial x_j}} - \overline{u_i\frac{\partial p}{\partial x_i}}. 
\label{eq:temp_balance1}
\end{equation}
The relative importance of the different terms in equation~\eqref{eq:temp_balance} is analysed in figure~\ref{fig:temp_balance},
for representative flow cases with wall heating and wall cooling. 
Figure~\ref{fig:temp_balance} shows that the mean temperature balance of variable-property flows is
qualitatively similar to what found in passive scalar convection, with mean conduction dominating the near-wall
region and turbulent convection dominating the overall balance away from the wall.
However, notable differences are the nonlinearity of the total heat flux (on account of the definition of $\mathcal{R}$) and the presence of additional terms which are small but not zero.
Indeed, very close to the wall we find small contribution of the fluctuating conduction term,
which however remains much smaller than the mean conduction.
We recall that in the mean momentum equation~\eqref{eq:mmb} 
the fluctuating diffusion term $\overline{\mu'\mathrm{d} u'/\mathrm{d} y}$ was neglected,
which yields a similar contribution to momentum balance as $\overline{\lambda'\mathrm{d} T'/\mathrm{d} y}$ 
yields to the temperature balance (not shown).
The dissipation remains negligible for all cases considered here, 
due to the small Mach number under scrutiny,
confirming that all flow cases can be regarded are representative of incompressible flow.
We also note that the total stress $\mathcal{R}(\eta)$ in equation~\eqref{eq:temp_balance1} is
indistinguishable from the sum of the components, indicating excellent statistical convergence of the results.

Based on the DNS data, we then assume $\Psi\approx0$ and $\overline{\lambda'\mathrm{d} \theta'/\mathrm{d} y}\approx0$ and, 
in analogy with what done for the turbulent shear stress, we model the turbulent heat flux 
by introducing a thermal eddy diffusivity, such that
\begin{equation}
	\widetilde{\theta''v''}\frac{(1-\eta)}{(1-\mathcal{R})} =-\alpha_T \frac{\mathrm{d}\widetilde{\theta}}{\mathrm{d}y}.
\end{equation}
Following the same procedure used for the mean momentum equation, we then determine the corresponding kernel function,
\begin{equation}
	h_{cp}=\left(\frac{1 + \alpha_T/\alpha}{1+\alpha_{T,cp}/\alpha}\right)\frac{(1-\eta)}{(1-\mathcal{R})}R N \frac {\mathrm{d}}{\mathrm{d} y} \left( \frac y{R^{1/2} N} \right),
\label{eq:TL_temp}
\end{equation}
where $\alpha=\lambda/(\rho c_p)$ is the thermal diffusivity coefficient.
Following~\citet{pirozzoli_23a}, we model the turbulent diffusivity as follows,
\begin{equation}
	\displaystyle\frac{\alpha_{T,cp}}{\alpha} = \frac{\left(\kappa_\theta y_{cp}^+\right)^3}{\left(\kappa_\theta y_{cp}^+\right)^2 + C_{v3}^2},\quad \displaystyle\frac{\alpha_T}{\alpha} = \frac{\left(\kappa_\theta y_{cp}^+\right)^3}{\left(\kappa_\theta y_{cp}^+\right)^2 + C_{v3}^2 + \beta(T_m/T_w)}
\label{eq:kernel_T}
\end{equation}
with constants $\kappa_\theta=0.459$, $C_{v3}= 10$.
As for the mean velocity, the dependency of the eddy thermal diffusivity 
on mean-to-wall temperature ratio is empirically accounted for by fitting the DNS data, to obtain
\begin{equation}
 \beta(T_m/T_w) = 
\begin{cases}
	(1-T_m/T_w)[141 - 507(T_m/T_w) + 608(T_m/T_w)^2], \quad  & T_m/T_w<1\\ 
	-28\log{(T_m/T_w)} + 1.6(1-T_m/T_w)^2, \quad & T_m/T_w>1.
\end{cases}
\label{eq:gfunc}
\end{equation}
To summarize, the kernel functions~\eqref{eq:TL} and~\eqref{eq:TL_temp} are rooted in the mean momentum balance and temperature balance equations, with eddy viscosity and
eddy conductivity to be specified. This is in our opinion more robust that
relying entirely on data-driven transformations~\citep{volpiani_20}. 
Nonetheless, we are not aware of any exact result in turbulence theory which does not include constants to be determined
from experience or simulation and the present case is no exception.
Here, the constants $\kappa$, $\kappa_{\theta}$, $C_{v1}$, $C_{v3}$ were determined once and for all for constant-property flow~\citep{pirozzoli_21,pirozzoli_23}.
The only added ingredients here are the functions $\varphi$ in equation~\eqref{eq:ffunc} and $\beta$ in~\eqref{eq:gfunc}, which account empirically for the dependency on the bulk-to-wall temperature ratio.
We note that both equations~\eqref{eq:ffunc} and \eqref{eq:gfunc} show different functional dependency for 
heating and cooling, which is aligned with empirical formulas for the Nusselt number and friction coefficient reported in the literature~\citep{petukhov_70,yeh_84}, 
featuring different coefficients or functions for the two cases.

\begin{figure}
 \begin{center}
 \includegraphics[scale=1,clip]{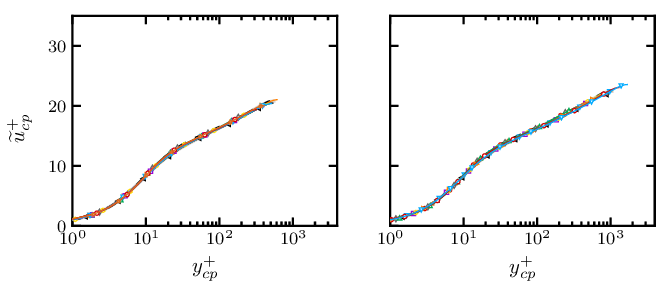}
  \put(-310,130){(\textit{a})}
  \put(-150,130){(\textit{b})}\\
 \includegraphics[scale=1,clip]{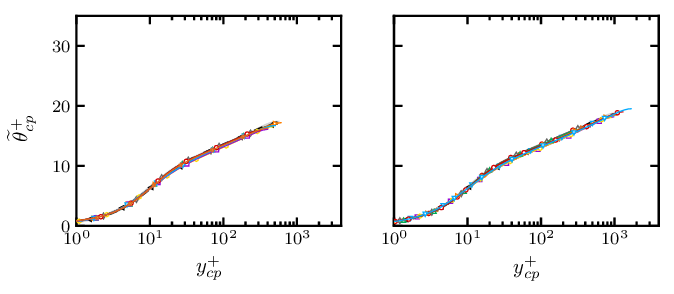}
  \put(-310,130){(\textit{c})}
  \put(-150,130){(\textit{d})}\\
	 \caption{Mean velocity (\textit{a},\textit{b}) and mean temperature (\textit{c},\textit{d}) profiles
	 transformed using equation~\eqref{eq:transf} with kernel functions~\eqref{eq:TL} and~\eqref{eq:kernel_T}, 
	 for L flow cases (\textit{a},\textit{c}) and H flow cases (\textit{b},\textit{d}).
	 Symbols indicate DNS data for different mean-to-wall
	 temperature ratios: 
	 $T_m /T_w = 0.4$ (left triangles), 
	 $T_m /T_w = 0.5, T_w=800K$ (downward triangles),
	 $T_m /T_w = 0.5, T_w=273.25K$ (right triangles), 
	 $T_m /T_w = 0.7$ (squares), 
	 $T_m /T_w = 0.8$ (hexagons), 
	 $T_m /T_w = 1.5$ (stars), 
	 $T_m /T_w = 2$   (circles), 
	 $T_m /T_w = 2.5$ (diamonds), 
	 $T_m /T_w = 3$ (upward triangles).
	 The gray solid lines indicate the reference mean velocity and temperature profiles of the constant-property case at $\Pran=0.72$,
	 obtained using the synthetic velocity profile of~\citet{musker_79} and the synthetic temperature profile of~\citet{pirozzoli_23a}. The dashed black lines indicate DNS of constant-property channel flow from~\citet{pirozzoli_16} at $\Pran=0.71$. 
	 }
  \label{fig:mean_transf}
 \end{center}
\end{figure}

In figure~\ref{fig:mean_transf}, we plot the transformed mean velocity and temperature profiles using the kernel functions~\eqref{eq:TL} and ~\eqref{eq:kernel_T}, 
and compare the results with the reference constant-property case.
The universality of the various distributions is quite remarkable, given the wide range
of variation of the flow properties which we are considering.
The accuracy of the velocity and temperature transformations 
also supports the validity of the assumptions made to derive
the kernel functions for the convolution integrals~\eqref{eq:transf}.
We point out that the coefficients inferred from DNS have been calibrated only for flow cases H, 
and they are successfully applied to lower or higher Reynolds number with similar accuracy, showing substantial 
independence from the Reynolds number.
These transformations allow us to define an equivalent channel height $h_{cp}$, which we
use to introduce an equivalent constant-property friction Reynolds number,
\begin{equation}
	\Rey_{\tau,cp} = \frac{h_{cp}}{\delta_{v}}, \quad h_{cp} = \int_0^hf_{cp}\mathrm{d}y. \label{eq:Retaucp}
\end{equation}
The equivalent channel height $h_{cp}$ is larger than $h$ for heating and smaller for cooling, leading to higher
or lower equivalent constant-property friction Reynolds number, respectively.
The definition given in equation~\eqref{eq:Retaucp}, can also be used to define an equivalent 
constant-property friction velocity, and an equivalent viscous length scale, namely
\begin{equation}
	\Rey_{\tau,cp} = \frac{u_{\tau,cp}h}{\overline{\nu}_w}, \quad u_{\tau,cp}=\frac{\overline{\nu}_w}{\delta_{v,cp}}, \quad \delta_{v,cp}  = \frac{h}{h_{cp}}\delta_v
\label{eq:retau_inc}
\end{equation}
Values of the equivalent constant-property Reynolds numbers are reported in table~\ref{tab:dataset}, 
which can be used as a guideline to interpret the instantaneous flow field in figure~\ref{fig:inst}, 
where flow cases with heating show finer eddies than for cooling because their effective Reynolds number is higher.

\section{Wall friction and heat transfer}

\begin{figure}
 \begin{center}
  \includegraphics[scale=1.0]{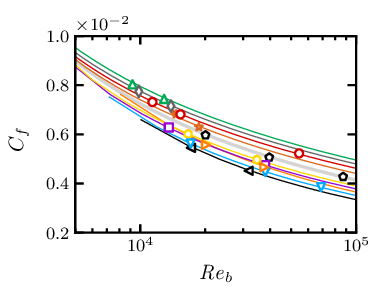}
  \includegraphics[scale=1.0]{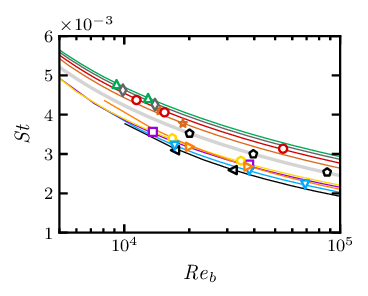}
  \put(-350,138){(\textit{a})}
  \put(-170,138){(\textit{b})}\\
  \includegraphics[scale=1.0]{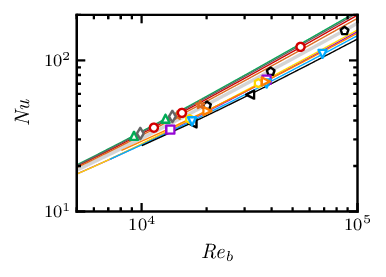}
  \includegraphics[scale=1.0]{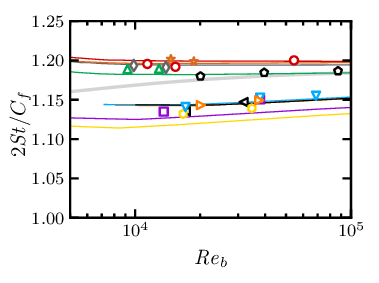}
  \put(-350,138){(\textit{c})}
  \put(-170,138){(\textit{d})}\\
  \caption{Friction coefficient (\textit{a}), Stanton number (\textit{b}), Nusselt number (\textit{c})
	 and Reynolds analogy factor (\textit{d}) as a function
	 of the bulk Reynolds number $\Rey_b=2h\rho_bu_b/\mu_m$.
	 Solid lines indicate predictions obtained by inverting the 
	 variable-property transformations~\eqref{eq:transf},
	 and symbols indicate DNS data for different mean-to-wall
	 temperature ratios, with matching colors: 
	 $T_m /T_w = 0.4$ (black left triangles), 
	 $T_m /T_w = 0.5, T_w=800K$ (orange downward triangle),
	 $T_m /T_w = 0.5, T_w=273.25K$ (blue right triangles), 
	 $T_m /T_w = 0.7$ (purple squares), 
	 $T_m /T_w = 0.8$ (gold hexagons), 
	 $T_m /T_w = 1.5$ (brown stars), 
	 $T_m /T_w = 2$   (red circles), 
	 $T_m /T_w = 2.5$ (dark gray diamonds), 
	 $T_m /T_w = 3$ (green upward triangles).
	 Black pentagons refer to DNS data of passive scalar in plane channel flow with $\Pran=0.71$ from~\citet{pirozzoli_16}.
	 }
  \label{fig:coeffs}
 \end{center}
\end{figure}

\begin{figure}
 \begin{center}
  \includegraphics[scale=1.0]{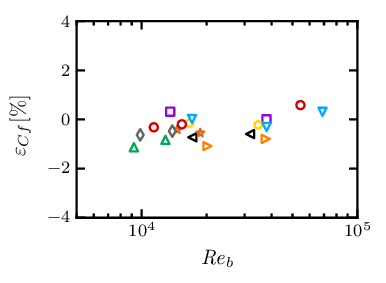}
  \includegraphics[scale=1.0]{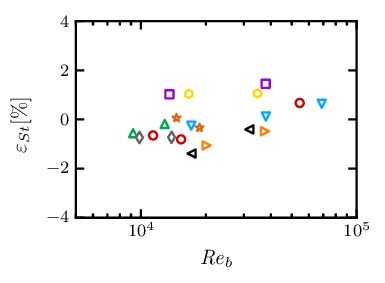}
  \put(-350,138){(\textit{a})}
  \put(-170,138){(\textit{b})}\\
	 \caption{Percent difference between DNS data and
	 predicted friction coefficient (\textit{a})
	 and Stanton number (\textit{b}) as a function
	 of the bulk Reynolds number $\Rey_b=2h\rho_bu_b/\mu_m$.
	 Symbols indicate DNS data for different mean-to-wall
	 temperature ratios: 
	 $T_m /T_w = 0.4$ (left triangles), 
	 $T_m /T_w = 0.5, T_w=800K$ (downward triangle),
	 $T_m /T_w = 0.5, T_w=273.25K$ (right triangles), 
	 $T_m /T_w = 0.7$ (squares), 
	 $T_m /T_w = 0.8$ (hexagons), 
	 $T_m /T_w = 1.5$ (stars), 
	 $T_m /T_w = 2$   (circles), 
	 $T_m /T_w = 2.5$ (gray diamonds), 
	 $T_m /T_w = 3$ (upward triangles).
	 }
  \label{fig:error}
 \end{center}
\end{figure}

The variable-property transformations developed in the previous section are very useful especially
as they enable the prediction of the heat transfer and friction coefficients. 
For that purpose, the only required inputs are the reference constant-property mean velocity and mean temperature profiles. As previously discussed, in the current work we consider the composite profiles 
developed in the work of~\citet{pirozzoli_24}.
However, different choices are possible, and one could for instance use 
the model for the mean velocity by~\citet{nagib_08}, and the model for the mean temperature
by~\citet{kader_81}, although the latter might result
in less accurate temperature profiles due to inconsistencies in the near-wall region.
Starting from those, application of the inverse of transformations~\eqref{eq:transf},
\begin{equation}
	y=\int_0^y \frac{1}{f_{cp}} \, {\mathrm d}y_{cp}, \quad
	\tilde{u}=\int_0^{u_{cp}} \frac{1}{g_{cp}} \, {\mathrm d}u_{cp}, \quad \tilde{\theta}=\int_0^{\theta_{cp}} \frac{1}{h_{cp}} \, {\mathrm d}\theta_{cp},
	\label{eq:transf_inv}
\end{equation} 
allows us to determine the actual variable-property profiles, 
for any given mean-to-wall temperature ratio and Reynolds number.
The key technical difficulty is that the kernel functions $f_{cp}$, $g_{cp}$, $h_{cp}$ depend on the 
actual temperature in the variable-property case, hence an iterative procedure is necessary,
as for compressible flow~\citep{kumar_22,hasan_24}.
The iterative procedure is presented in~\ref{alg:iter} and it can be summarized as follows,

\begin{enumerate}
  \item Generate the constant-property profiles for a target friction Reynolds number $\Rey_{\tau,cp}$		
  \item Initialize kernel functions $f_{cp}$, $g_{cp}$, $h_{cp}$ using the constant-property temperature profile
  \item Calculate the backward convolution integrals~\eqref{eq:transf_inv} to find $y$, $u$, $\theta$
  \item Update the kernels $f_{cp}$, $g_{cp}$, $h_{cp}$ using the newly calculated temperature $\theta$
  \item Calculate the friction coefficient, Stanton number and $\Rey_{\tau,cp}$ 
  \item Update the constant-property profiles based on the updated $\Rey_{\tau,cp}$
\end{enumerate}

Note that two nested loops are required
for this iterative algorithm. This is because the constant-property profiles are also re-calculated at each step
in order to converge towards the target friction Reynolds number $\Rey_{\tau,cp}$.

\begin{algorithm}
	\caption{Inverse variable-property transformation: $\varepsilon_{\Rey_\tau}$ and $\varepsilon$ are the tolerances for the iterative algorithm, which in our case are set to $10^{-9}$ and $10^{-10}$, respectively.}
\begin{algorithmic}[1] 
 \Statex \textbf{Initialization:}\\ 
	Generate constant-property profiles for a prescribed target value of ${\Rey_{\tau,cp}}$: $u_{cp}^n$, $\theta_{cp}^n$
	\State Set: $u^n=u_{cp}^n, \theta^n=\theta_{cp}^n$, Calculate: $f_{cp}^n, g_{cp}^n, h_{cp}^n, C_f^{n}, \St^{n}$ using $\theta^n$ for a given $T_m/T_w$  
	\vspace{0.5cm}
	\Statex \textbf{while} {$(|\Rey_{\tau,cp}^{n}-{\Rey_{\tau,cp}}|<\varepsilon_{\Rey_\tau})$} \textbf{do}
	\vspace{0.5cm}
	\Statex\quad \textbf{while} {$(|C_f^{n+1}-C_f^n|<\varepsilon)\, \text{and}\, (|\St^{n+1}-\St^n|<\varepsilon)$} \textbf{do}
	\vspace{0.5cm}
	\State\quad\quad $y^{n+1} = \displaystyle\int \frac{1}{f_{cp}^n}\mathrm{d}y_{cp}^n$, $\quad \tilde{u}^{n+1} = \displaystyle\int \frac{1}{g_{cp}^n}\mathrm{d}u_{cp}^n,$ $\quad\tilde{\theta}^{n+1} = \displaystyle\int \frac{1}{h_{cp}^n}\mathrm{d}\theta_{cp}^n$
	\vspace{0.5cm}
	\State\quad\quad Update kernels: $f_{cp}^{n+1}, g_{cp}^{n+1}, h_{cp}^{n+1}$, using equations~\eqref{eq:TL}--\eqref{eq:TL_temp}
	\vspace{0.5cm}
	\State\quad\quad Update coefficients and Reynolds: $C_f^{n+1}, \St^{n+1}, \Rey_{\tau,cp}^{n+1}$ using equation~\eqref{eq:Retaucp}
	\vspace{0.5cm}
	\State\quad\quad Update constant-property profiles $u_{cp}^{n+1}$, $\theta_{cp}^{n+1}$ at $\Rey_{\tau,cp}^{n+1}$
    \Statex\quad\textbf{end while}
    \Statex\textbf{end while}
\end{algorithmic}
\label{alg:iter}
\end{algorithm}

Figure~\ref{fig:coeffs} shows the resulting friction coefficient $C_f=2\tau_w/(\rho_bu_b^2)$, and 
the Stanton number $\St = q_w/[\rho_b C_p u_b\left(T_w-T_m\right)]$.
Whereas the data for the constant-property case (black pentagons) fall on top of the corresponding 
theoretical curves (light gray), we find significant deviations thereof in cases with properties variations.
In particular, we note that cases with heated wall yield reduced friction and heat flux, whereas
cases with cooled wall yield an increase of momentum and heat transfer, with a scatter around the constant-property 
case of $\pm 25 \%$ for both the friction coefficient and the Stanton number.
We also report the heat transfer in terms of Nusselt number $\Nu = \Rey_b \St \Pran$, although 
this representation tends to hide differences within a few percent, thus the Stanton number should be preferred 
for accurate evaluation of theories.
Theoretical predictions relying on use of the variable-property transformations~\eqref{eq:transf} are 
shown in the figure with solid line of matching colors, and of course those are not universal as well.
Notably, figure~\ref{fig:coeffs} shows that the resulting predictions match the DNS data to within 1-2\% accuracy
for all cases, both for the friction and heat transfer coefficients, 
as shown quantitatively in figure~\ref{fig:error}.
We further find that the analogy between momentum and heat transfer holds also in the case of 
variable-property flows, as the Reynolds analogy factor stays close to the constant property case, 
although this information alone is obviously not sufficient to recover the heat transfer and friction coefficients 
from the constant-property case. Note that the Reynolds analogy factor is not unity even in the constant-property case
because $\Pran=0.72$.

\section{Conclusions}

Currently, predicting heat transfer through forced convection in real fluids heavily depends 
on fitting experimental data obtained decades ago, leading to uncertainties of up to 20-30\%. 
This significant variability is clearly reflected in the current DNS data.
To address this uncertainty, we have developed a robust framework for estimating momentum 
and heat transfer coefficients. Our approach is grounded in the first principles of momentum 
and energy balance rather than empirical methods, offering the advantages of accuracy and generalizability. 
Similar to approaches used in high-speed turbulent boundary layers, our method relies 
on transformation kernels for velocity and temperature distributions.

Preliminary tests indicate that transformation kernels informed by DNS data 
can generate velocity and temperature distributions with excellent universality 
compared to the constant-property case. Evaluating momentum and heat transfer coefficients 
involves integrating the estimated velocity and temperature profiles obtained through the 
backward application of these transformation kernels, requiring an iterative procedure.
Our results indicate that the method can accurately predict heat transfer and friction 
coefficients within 1-2\% compared to DNS data. 
Additionally, the developed method can determine mean temperature and velocity profiles alone, 
providing valuable information for establishing wall functions in simulations employing wall-modeled approaches.

We also acknowledge that cooling ducts in practical applications often feature 
rough walls rather than smooth ones~\citep{chung_21,zhong_23,demaio_23}. 
This raises questions about the applicability of the current transformations 
to complex surface patterns. 
In such cases, the effects of density and viscosity 
variations lead to a much more intricate flow physics compared to constant-property flows. 
Several mechanisms and parameters remain to be studied, including the precise mechanisms 
responsible for friction and heat transfer variation, the impact on turbulence 
length scales, the dependency of Prandtl number and heat capacity on the temperature, and the influence of the working fluid. 
We plan to investigate these aspects in future studies.

We believe that the proposed approach could have important implications in closely related fields, as mixed and natural convection. In principle, we see no reason why the same approach shouldn't be applicable
these flow problems, as the transformations are rooted in the mean momentum and thermal balance, which are universal.\\

\noindent{\bf Data availability}\\
DNS data are available at \url{http://newton.dma.uniroma1.it/}. Model coefficients can be generated at \url{http://www.thermoturb.com}.\\

\noindent{\bf Acknowledgements}\\
We acknowledge CHRONOS for awarding us access to Piz Daint, at the Swiss National Supercomputing Centre (CSCS), Switzerland.
We also acknowledge EuroHPC for access to LEONARDO based at CINECA, Casalecchio di Reno, Italy.

\appendix
\section{}\label{appA}
We solve the fully compressible Navier--Stokes equations for a perfect heat-conducting gas,
\begin{subequations}
 \begin{align}
  \frac{\partial \rho}{\partial t}   +
  \frac{\partial\rho u_i}{\partial x_i}  & = 0 ,
  \label{eq:mass}\\
  \frac{\partial \rho u_i}{\partial t}  +
  \frac{\partial\rho u_i u_j}{\partial x_j} &= -
  \frac{\partial p}{\partial x_i}           +
  \frac{\partial\sigma_{ij}}{\partial x_j}  +
  f \delta_{i1} ,
  \label{eq:momentum}\\
  \frac{\partial \rho s}{\partial t} +
	 \frac{\partial\rho u_j s}{\partial x_j} &= \frac{1}{T}\left(-
  \frac{\partial q_j}{\partial x_j}       +
	 \sigma_{ij}\frac{\partial u_i}{\partial x_j}\right)   +
  Q,
  \label{eq:entropy}
 \end{align}
\end{subequations}
where $u_i$, $i=1,2,3$, is the velocity component in the i-th direction, $\rho$ the density,
$p$ the pressure, $s=c_v\log(p\rho^{-\gamma})$ the entropy per unit mass, and $\gamma=c_p/c_v=1.4$ is the specific heat ratio.
The components of the heat flux vector $q_j$ and of the viscous stress tensor $\sigma_{ij}$ are
\begin{equation}
 \sigma_{ij}=\mu\left(\frac{\partial u_i}{\partial x_j} + \frac{\partial u_j}{\partial x_i} -\frac{2}{3} \frac{\partial u_k}{\partial x_k} \delta_{ij}\right),
\end{equation}
\begin{equation}
 q_j=-k\frac{\partial T}{\partial x_j},
\label{eq:heat_vec}
\end{equation}
where the dependence of the viscosity coefficient on temperature is accounted for through Sutherland's law
and $k=c_p\mu/\Pran$ is the thermal conductivity, with $\Pran=0.72$.
Use of the entropy equation to replace the energy equations is here dictated from the 
possibility to relax the acoustic time step limitation with semi-implicit time-stepping~\citet{modesti_18a},
thus yielding a computational efficiency comparable to a variable-property incompressible solver.

The forcing term $f$ in equation~\eqref{eq:momentum} is evaluated at each time step in order to discretely enforce constant
mass flow rate in time.
Similarly, a uniform bulk heating/cooling term is added to the entropy equation to make the mixed mean temperature to be exactly constant in time.
Since we solve for the entropy equation, this is achieved by correcting the local temperature at each grid point and each Runge--Kutta sub-step as follows
\begin{equation}
	T(x,y,z,t) \rightarrow T(x,y,z,t) - T_m(t) +  T_m^*,
\end{equation}
where $T_m(t)$ is the mixed mean temperature before the correction,
and $T_m^*$ is the target value. The updated temperature value is then used to re-evaluate the entropy at the current time.\\

\noindent{\bf Declaration of interests}\\
The authors report no conflict of interest.

\addcontentsline{toc}{chapter}{Bibliography}
\bibliographystyle{jfm}
\bibliography{references} 
\end{document}